%% file: main.tex
\documentclass[conference]{IEEEtran}
\IEEEoverridecommandlockouts
\usepackage[bottom=1.02in,top=0.71in,left=0.64in,right=0.64in]{geometry}
\usepackage{amsmath}
\usepackage{amsfonts}
\usepackage{amssymb}
\usepackage[ruled,vlined]{algorithm2e}
\usepackage{caption}
\usepackage{mathtools}
\usepackage{amsthm}
\usepackage{algpseudocode}
\usepackage{booktabs}
\usepackage{subfigure}
\usepackage{tabulary}
\usepackage{footnote}
\usepackage[export]{adjustbox}
\usepackage[justification=centering]{caption}
\usepackage{comment}
\usepackage{cases}
\usepackage{graphicx}
\usepackage{cite}
\usepackage{multicol}
\usepackage{xcolor}
\usepackage{graphicx}
\graphicspath{{./}{figures/}}

\usepackage{stfloats}
\usepackage{cuted}
\usepackage{lipsum}

\usepackage{dsfont}
\usepackage{optidef}

\title{EPFL-REMNet: Efficient Personalized Federated Digital Twin Towards 6G Heterogeneous Radio Environment}

\author{
    \IEEEauthorblockN{
        Peide Li\IEEEauthorrefmark{1}, 
        Liu Cao\IEEEauthorrefmark{1},
        Lyutianyang Zhang\IEEEauthorrefmark{2}, Dongyu Wei\IEEEauthorrefmark{3}, Ye Hu\IEEEauthorrefmark{4},
        Qipeng Xie \IEEEauthorrefmark{5}\\
    }
    \IEEEauthorblockA{
        \IEEEauthorrefmark{1}Department of Electronic Information Engineering, City University of Hong Kong (Dongguan), Dongguan, China\\
        \IEEEauthorrefmark{2}The School of Microelectronics and Communication Engineering, Chongqing University, Chongqing, China\\
        \IEEEauthorrefmark{3}Department of Electrical and Computer Engineering, University of Miami, FL, USA\\
        \IEEEauthorrefmark{4}Department of Industrial and Systems Engineering, University of Miami, FL, USA\\
        \IEEEauthorrefmark{5}Information Hub, Hong Kong University of Science and Technology (Guangzhou), Guangzhou, China\\
        Emails: \{peide.li, liu.cao\}@cityu-dg.edu.cn, zhanglyutianyang@cqu.edu.cn, \{dongyu.wei,yehu\}@miami.edu, \\qxieaf@connect.ust.hk
        \vspace{-0.5cm}
    }

\thanks{This work was supported by the Youth Innovation Talent Project of Guangdong Provincial Universities (Grant No. 2025KQNCX17).}
}

\begin{document}

\maketitle
\thispagestyle{empty}
\begin{abstract}
Radio Environment Map (REM) is transitioning from 5G homogeneous environments to B5G/6G heterogeneous landscapes. However, standard Federated Learning (FL), a natural fit for this distributed task, struggles with performance degradation in accuracy and communication efficiency under the non-independent and identically distributed (Non-IID) data conditions inherent to these new environments. This paper proposes EPFL-REMNet, an efficient personalized federated framework for constructing a high-fidelity digital twin of the 6G heterogeneous radio environment. The proposed EPFL-REMNet employs a ``shared backbone + lightweight personalized head” model, where only the compressed shared backbone is transmitted between the server and clients, while each client's personalized head is maintained locally. We tested EPFL-REMNet by constructing three distinct Non-IID scenarios (light, medium, and heavy) based on radio environment complexity, with data geographically partitioned across 90 clients. Experimental results demonstrate that EPFL-REMNet simultaneously achieves higher digital twin fidelity (accuracy) and lower uplink overhead across all Non-IID settings compared to standard FedAvg and recent state-of-the-art methods. Particularly, it significantly reduces performance disparities across datasets and improves local map accuracy for long-tail clients, enhancing the overall integrity of digital twin.
\end{abstract}


\input{introduction}

\input{sys_model}

\input{simulation}

\input{conclusion}

\bibliographystyle{IEEEtran}
\bibliography{ref}
\end{document}

%% file: introduction.tex
\section{Introduction}
\label{sec:intro}
The construction of a real-time, high-fidelity Radio Environment Map (REM) is emerging as a cornerstone technology for the intelligent operation and optimization of 5G-Advanced and future 6G networks. A REM functions as a digital twin of the radio environment, mapping key performance indicators like received power and path loss to replace costly ray-tracing simulations with a dynamic, data-driven model \cite{lee2024scalable}. Recent academic and industrial efforts, such as challenges focused on path loss estimation, have helped standardize datasets and benchmarks, fostering collaborative progress \cite{yapar2024overview}. In practice, the measurement data required to build this digital twin is naturally distributed across end-user devices and edge servers. This distribution, coupled with privacy regulations and uplink bandwidth constraints, makes centralized training untenable. Federated Learning (FL), with its principle of ``training locally while building a shared model," offers a compelling paradigm for this task, enabling the collaborative construction of a network-wide digital twin from decentralized data sources \cite{lee2024federated}.

Despite its promise, deploying FL for REMs immediately confronts two formidable challenges. The first is {\em extreme statistical heterogeneity}. 6G wireless signal propagation is strongly correlated with geography; diverse urban topographies: building densities, and material compositions create vast differences in data distributions across clients \cite{fine2021tuning, geographical2018clustering}. This location-dependent, structural Non-IID nature of the data means a single global model often fails to generalize to specific local environments, particularly for ``long-tail" clients with unique propagation characteristics. The second challenge is the {\em severe communication bottleneck}. In edge-centric wireless networks, the uplink is typically the most constrained resource. The iterative optimization process of FL requires frequent exchanges of high-dimensional model parameters or gradients, imposing significant burdens on network bandwidth and operational costs, thereby hindering large-scale, continuous training \cite{spiridonoff2021local}.

Existing research has addressed the above issues in two separate directions: Personalized Federated Learning (PFL) is used to tackle the heterogeneity, for example, splitting models into a shared backbone and local heads (e.g., FedRep \cite{collins2021fedrep}), applying personalized regularization (e.g., Ditto \cite{li2021ditto}), or using local normalization layers to mitigate feature shift (e.g., FedBN \cite{li2021fedbn}). While being effective from the adaptation perspective, these methods do not inherently reduce communication costs. Besides, the communication-efficient FL focuses on minimizing data transmission through techniques such as sparsification with error feedback \cite{richtarik2021ef21}, quantization \cite{sun2024low}, and reduced communication frequency \cite{gorbunov2021marina}. However, under strong Non-IID conditions, aggressive compression can discard fine-grained gradient information crucial for local adaptation, leading to instability and slower convergence. This suggests that personalization and communication efficiency are not independent problems. A robust personalization strategy can make the global aggregation task more amenable to compression, while a well-calibrated compression scheme can, in turn, regularize the shared model to benefit personalization \cite{LC2025IoTJ,hu2024multi,cao2022resource,yin2022routing,zhang2025cross}.


Motivated by the aforementioned issues, we propose an Efficient PFL(EPFL)- REMNet, a framework that synergistically integrates both concepts. The key contributions are summarized as follows:
\begin{itemize}
    \item \textbf{A Novel Co-Design Framework:} We deeply couple a PFL architecture (``shared backbone + local personalized heads") with a hybrid communication compression pipeline (Top-K sparsification, error feedback, 8-bit quantization, and periodic updates). This design achieves both bandwidth savings and stable convergence by ensuring a proper division of labor between transferability and adaptability.
    \item \textbf{A Comprehensive Evaluation Methodology:} We construct light, medium, and heavy Non-IID scenarios to mirror real-world data heterogeneity. Beyond standard RMSE/MAE metrics, we introduce fairness and robustness indicators to quantify both model utility and engineering trustworthiness.
    \item \textbf{State-of-the-Art Performance:} Across all heterogeneity levels, EPFL-REMNet demonstrates significant advantages over standard FedAvg \cite{mcmahan2017communication} and recent baselines like FedFly \cite{varyam2022towards} and PFedgb \cite{dmitriev2021pfedgp}. Compared to uncompressed PFL, it reduces cumulative communication overhead by over 97\% with negligible accuracy loss, while simultaneously improving performance for long-tail clients.
    \item \textbf{A Practical Multi-Objective Optimization Paradigm:} We frame the construction of the REM-based digital twin as a trade-off between accuracy and communication cost, and we map the Pareto frontier. This provides an actionable process for hyperparameter selection, enabling deployment configurations that meet specific operational constraints, such as minimizing communication cost for a target digital twin fidelity.
\end{itemize}

The rest of this paper is organized as follows: Sec \ref{sec:sys_arc} details the system architecture and problem formulation of the EPFL-REMNet framework while Sec \ref{sec:algo} provides the corresponding federated optimization algorithm. The experimental design used to rigorously evaluate EPFL-REMNet, including the dataset, heterogeneity scenario construction, evaluation metrics, and baseline methods are detailed in Sec \ref{sec:sim}. Finally, section \ref{sec:con} draws the conclusions for this paper.

%% file: sys_model.tex
\section{The EPFL-REMNet Framework}
\label{sec:sys_arc}

\subsection{System Architecture}

We consider a 6G heterogeneous radio environment scenario (including significantly distinct
light, medium, and heavy radio environment areas that are Non-IID) with $N$ clients (e.g., edge servers representing distinct geographical areas) and $M$ base stations (BSs). Each client $i = \{1,2,...,N\}$ possesses a local dataset $\mathcal{D}_i$.

\begin{equation}
\mathcal{D}_i = \{(\boldsymbol{X}_j, \boldsymbol{y}_j)\}_{j=1}^{|\mathcal{D}_i|},
\label{eq:local_dataset}
\end{equation}

\noindent where the input feature vector $\boldsymbol{X}_j \in \mathbb{R}^{2+P}$ comprises 2-dimension spatial coordinates and $P$ environmental auxiliary features. The output label $\boldsymbol{y}_j \in \mathbb{R}^{M}$ represents the signal strength from $M$ BSs:

\begin{equation}
\boldsymbol{y}_j = \left[s_1, s_2, ..., s_M\right]^T.
\label{eq:label_vector}
\end{equation}

To handle data heterogeneity, each client performs local preprocessing. Spatial coordinates are normalized to the range $[0,1]$ using min--max scaling. The signal strength labels are standardized using the client's local mean $\mu_i$ and standard deviation $\sigma_i$ via z-score normalization:

\begin{equation}
\hat{\boldsymbol{y}} = \frac{\boldsymbol{y} - \boldsymbol{\mu}_i}{\sigma_i}.
\label{eq:zscore}
\end{equation}

\begin{figure}[htbp]
\centerline{\includegraphics[width=\columnwidth]{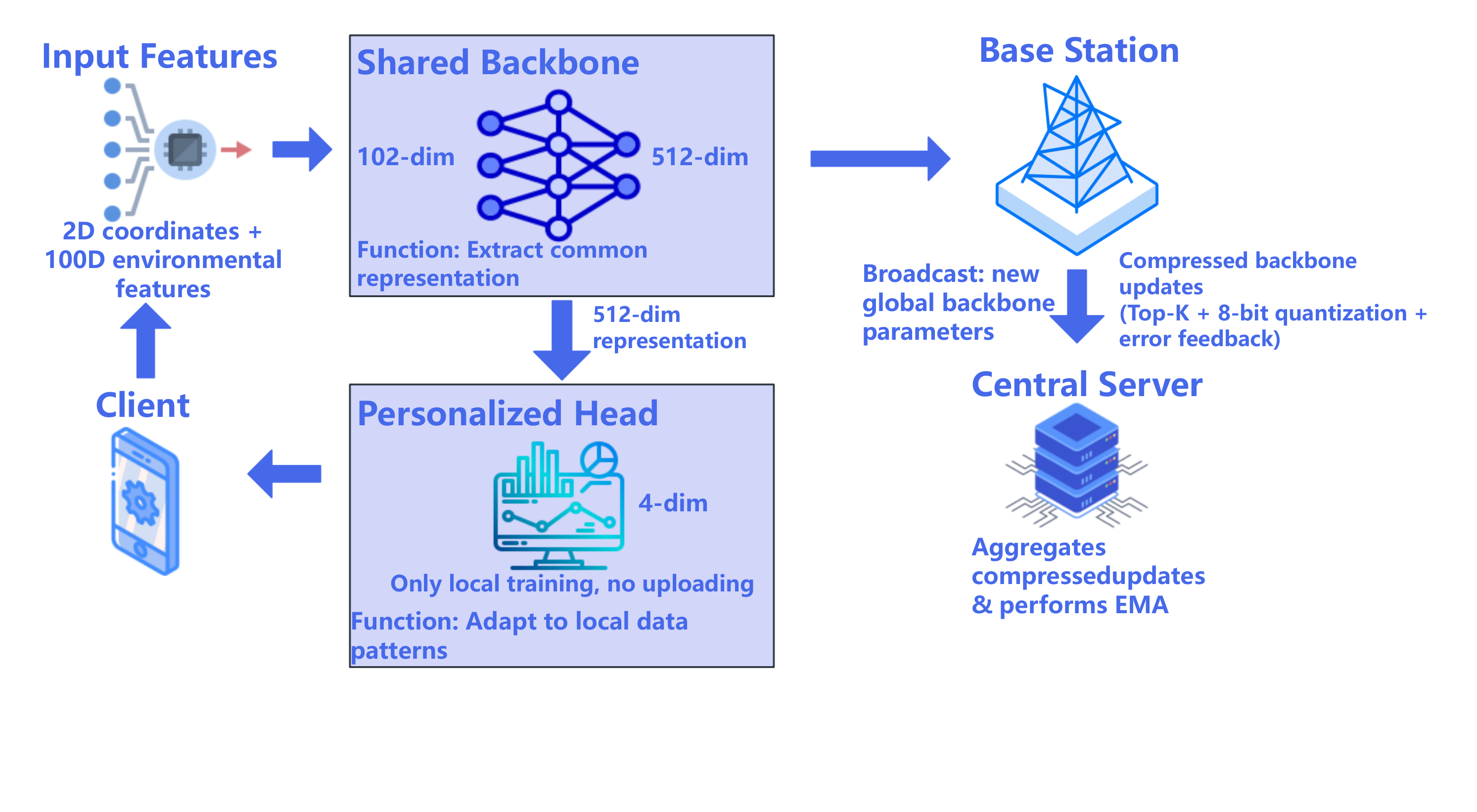}}
\vspace{-0.7cm}
\caption{EPFL-REMNet System Architecture.}
\label{fig:system_architecture}
\end{figure}

As Fig.~\ref{fig:system_architecture} shows, the proposed EPFL-REMNet's architecture embodies the principle of decoupling generalizable knowledge from local specificities.

\textbf{Shared Backbone Network ($f_{\boldsymbol{\theta}}: \mathbb{R}^{2+P} \to \mathbb{R}^{512}$):}
This component is responsible for extracting high-level, transferable representations from the input features and is shared among all clients. It consists of a three-layer Multi-Layer Perceptron (MLP) that maps the input to a 512-dimensional latent representation $\boldsymbol{z} \in \mathbb{R}^{512}$. Its structure is

\begin{equation}
z = f_{\boldsymbol{\theta}}(x) = \mathcal{N}_3\!\Big( \mathcal{A}\!\big(\boldsymbol{W}_3 \cdot \mathcal{N}_2\!( \mathcal{A}(\boldsymbol{W}_2 \cdot \mathcal{N}_1\!( \mathcal{A}(\boldsymbol{W}_1 x)) ))\big)\Big),
\label{eq:backbone}
\end{equation}

where $\boldsymbol{W}_1, \boldsymbol{W}_2, \boldsymbol{W}_3$ are weight matrices, $\mathcal{A}(\cdot)$ is the SiLU activation function, and $\mathcal{N}_n(\cdot)$ denotes LayerNorm of the $n$th layer. The use of LayerNorm is crucial for stabilizing the training process, especially under FL scenarios with heterogeneous data distributions.

\textbf{Personalized Head Network ($g_{\boldsymbol{\phi}_i}: \mathbb{R}^{512} \to \mathbb{R}^{M}$):} This lightweight network takes the latent representation $\boldsymbol{z}$ from the backbone and maps it to the final $M$-dimension signal strength estimation, which constitutes a data point within the local REM. Each client $i$ exclusively trains and maintains its own head ${g}_{\boldsymbol{\phi}_i}$, allowing it to specialize in the unique radio characteristics of its local environment, thereby ensuring a more accurate local representation within the global digital twin. We consider a two-head architecture:

\begin{itemize}
    \item \textbf{Single-layer head:} A simple linear layer, 
    \begin{equation}
        {g}_{\boldsymbol{\phi}_i}(\boldsymbol{z}) = \boldsymbol{W}_{\text{out}}^{(i)} \boldsymbol{z} + \boldsymbol{b}_{\text{out}}^{(i)}
    \end{equation}
    \item \textbf{Two-layer head:} An MLP with one hidden layer for greater non-linear capacity, 
    \begin{equation}
        {g}_{\boldsymbol{\phi}_i}(\boldsymbol{z}) = \boldsymbol{W}_2^{(i)} \sigma\left( \boldsymbol{W}_1^{(i)} \boldsymbol{D}(\boldsymbol{z}) \right) + \boldsymbol{b}_2^{(i)}
    \end{equation}
    Where $\boldsymbol{D}(\cdot)$ denotes the dropout operation. During training, it randomly drops out the neuron outputs of part of the input feature $\boldsymbol{z}$ to suppress model overfitting.
\end{itemize}

\subsection{Problem Formulation}

The system's objective is to minimize the weighted average of a loss function over all $N$ clients, which achieves the high-fidelity digital twin per BS radio environment. The model is decomposed into a shared backbone network $f_{\boldsymbol{\theta}}$ with parameters $\boldsymbol{\theta}$ and client-specific personalized head networks $g_{\boldsymbol{\phi}_i}$ with parameters $\boldsymbol{\phi}_i$ of client $i$. The global optimization problem is

\begin{equation}
\min_{\boldsymbol{\theta}, \{\boldsymbol{\phi}_i\}_{i=1}^N} 
\sum_{i=1}^{N} \frac{|\mathcal{D}_i|}{\sum_{j=1}^N |\mathcal{D}_j|}
\;\mathbb{E}_{(x,y) \sim \mathcal{D}_i} \!\left[ \mathcal{L}\!\big(g_{\boldsymbol{\phi}_i}(f_{\boldsymbol{\theta}}(x)), \hat{\boldsymbol{y}}\big) \right],
\label{eq:global_obj}
\end{equation}

where $\mathcal{L}(\cdot)$ is the Huber loss, chosen for its robustness to outliers by combining the advantages of Mean Squared Error (MSE) and Mean Absolute Error (MAE).

\section{Federated Optimization and Communication Method}
\label{sec:algo}
The training process of EPFL-REMNet is governed by a carefully designed protocol that co-optimizes for accuracy and communication efficiency, 

In a communication round $t$, the process is as follows:
\begin{enumerate}
    \item \textbf{Server Broadcast:} The server sends the current global backbone parameters $\boldsymbol{\theta}^t$ to a selected subset of clients.
    \item \textbf{Local Training:} Each participating client $i$ synchronizes its local backbone to $\boldsymbol{\theta}^t$. It then performs $E$ epochs of local training on its full model (backbone and head) using its local dataset $\mathcal{D}_i$. The personalized head parameters $\boldsymbol{\phi}_i$ are updated and stored locally, while the update to the backbone, $\boldsymbol{\Delta}_i^t$, is computed.
    \item \textbf{Update Compression:} To minimize uplink communication, the update $\boldsymbol{\Delta}_i^t$ undergoes a three-stage compression pipeline \cite{11096953}:
    \begin{itemize}
        \item \textbf{Error Feedback and Accumulation:} The current update $\boldsymbol{\Delta}_i^t$ is adjusted by incorporating the \emph{sparsification error} $\boldsymbol{e}_i^{t-1}$, 
        which represents the residual information from the previous round that was not transmitted due to sparsification. 
        Formally: 
        \begin{equation}
        \boldsymbol{u}_i^t = \boldsymbol{\Delta}_i^t + \boldsymbol{e}_i^{t-1}.
        \end{equation}
        This mechanism compensates for the information loss introduced by sparsification and helps maintain stable convergence.
        \item \textbf{Top-K Sparsification:} The client retains only the $K$ elements of $\boldsymbol{u}_i^t$ with the largest absolute values and sets the remaining ones to zero. 
        The new sparsification error is defined as
        \begin{equation}
        \boldsymbol{e}_i^t = \boldsymbol{u}_i^t - \mathcal{T}_K(\boldsymbol{u}_i^t, K),
        \end{equation}
        where $\mathcal{T}_K(\cdot, K)$ denotes the operation that preserves the $K$ largest-magnitude entries and zeroes out the rest. 
        This residual $\boldsymbol{e}_i^t$ is cached for the next round.
        \item \textbf{8-bit Symmetric Quantization:} The non-zero elements of the sparse vector are quantized to 8-bit integers. The client transmits only the indices of these elements, their 8-bit values, and a single scaling factor for reconstruction.
    \end{itemize}
    \item \textbf{Periodic Synchronization:} The compression and upload steps are performed only once every $R$ rounds. In the intermediate $R-1$ rounds, clients train locally without communicating with the server, drastically reducing the total number of communication rounds.
    \item \textbf{Server Aggregation:} The server collects the compressed updates, de-quantizes and decodes them, and aggregates them via averaging to update the global backbone model to $\boldsymbol{\theta}^{t+1}$. Optionally, the server can maintain an Exponential Moving Average (EMA) of the global parameters to further enhance training stability.
\end{enumerate}

In summary, the EPFL-REMNet training algorithm as outlined in Algorithm \ref{alg:pflremnet} for a single round.

\begin{algorithm}[t]
\footnotesize
\caption{EPFL-REMNet Training Algorithm}
\label{alg:pflremnet}
\SetAlgoLined
\KwIn{Global round $t$, global shared feature parameters $\boldsymbol{\theta}^t$, client sampling set $S_t$, local epochs $E$, communication period $R$}
\textbf{Server executes:}\\
Broadcast $\boldsymbol{\theta}^t$ to all clients $i \in S_t$\;
Initialize empty set of updates $\mathcal{U} = \emptyset$\;
\For{client $i \in S_t$ in parallel}{
    $\boldsymbol{\Delta}_i^{t} \leftarrow \text{ClientUpdate}(i, \boldsymbol{\theta}^t)$\;
    \If{$\boldsymbol{\Delta}_i^{t}$ is not null}{
        Add $\boldsymbol{\Delta}_i^{t}$ to $\mathcal{U}$\;
    }
}
$\boldsymbol{\theta}^{t+1} \leftarrow \boldsymbol{\theta}^t + \frac{1}{|\mathcal{U}|} \sum_{\boldsymbol{\Delta} \in \mathcal{U}} \boldsymbol{\Delta}$\;
\textbf{ClientUpdate($i, \boldsymbol{\theta}^t$):}\\
$\boldsymbol{\theta}_i \leftarrow \boldsymbol{\theta}^t$\;
\For{$e=1$ to $E$}{
    Train $(f_{\boldsymbol{\theta}_i}, g_{\boldsymbol{\phi}_i})$ on local data $\mathcal{D}_i$\;
}
$\boldsymbol{\Delta}_i^t \leftarrow \boldsymbol{\theta}_i - \boldsymbol{\theta}^t$\;
\If{$t \pmod R == 0$}{
    $\boldsymbol{u}_i^t \leftarrow \boldsymbol{\Delta}_i^t + \boldsymbol{e}_i^{t-1}$ \tcp*{Error feedback}
    $\tilde{\boldsymbol{u}}_i^t \leftarrow \mathcal{T}_K(\boldsymbol{u}_i^t, K)$ \tcp*{Sparsification}
    $\boldsymbol{e}_i^t \leftarrow \boldsymbol{u}_i^t - \tilde{\boldsymbol{u}}_i^t$ \tcp*{Cache error}
    $\hat{\boldsymbol{u}}_i^t \leftarrow \text{Quantize}(\tilde{\boldsymbol{u}}_i^t)$ \tcp*{Quantization}
    \Return $\text{Decode}(\hat{\boldsymbol{u}}_i^t)$ to server\;
}
\Else{
    \Return null\;
}
\end{algorithm}

%% file: simulation.tex
\section{Experimental Evaluation}
\label{sec:sim}

\begin{table*}[htbp]
\caption{Overall Performance Comparison Across Three Non-IID Scenarios.}
\label{tab:main_results}
\centering
\footnotesize
\resizebox{.65\textwidth}{!}{\begin{tabular}{@{}llcccc@{}}
\toprule
\textbf{Scenario} & \textbf{Method} & \textbf{RMSE (micro) $\downarrow$} & \textbf{RMSE (macro) $\downarrow$} & \textbf{MAE (macro) $\downarrow$} & \textbf{Communication (MB) $\downarrow$} \\
\midrule
\textbf{Light} & FedAvg & 6.66 & 6.63 & 5.22 & 86689.5 \\
& FedFly & 7.73 & 9.94 & 8.21 & 182111.6 \\
& PFedgb & 7.51 & 7.37 & 5.89 & 20610.4 \\
& \textbf{EPFL-REMNet} & \textbf{3.77} & \textbf{3.75} & \textbf{1.83} & \textbf{13437.6} \\
\midrule
\textbf{Medium} & FedAvg & 4.10 & 3.96 & 3.20 & 86689.5 \\
& FedFly & 4.71 & 4.71 & 3.77 & 182111.6 \\
& PFedgb & 5.16 & 5.03 & 4.02 & 20610.4 \\
& \textbf{EPFL-REMNet} & \textbf{1.37} & \textbf{1.33} & \textbf{0.88} & \textbf{7043.2} \\
\midrule
\textbf{Heavy} & FedAvg & 6.56 & 6.27 & 4.91 & 86689.5 \\
& FedFly & 4.57 & 4.58 & 3.88 & 182111.6 \\
& PFedgb & 5.67 & 5.63 & 4.76 & 20610.4 \\
& \textbf{EPFL-REMNet} & \textbf{2.07} & \textbf{2.02} & \textbf{1.40} & \textbf{21630.7} \\
\bottomrule
\end{tabular}}
\end{table*}

\subsection{Experimental Setup}
\textbf{Dataset and Preprocessing:} We use the public RadioMapSeer dataset \cite{dataset2021pathloss}, which provides rasterized path loss and received power maps for multiple base stations. We developed a custom script to process these maps into a unified format. For each data point, we extract the normalized 2D coordinates and the signal strengths for four base stations (1, 2, 3, 4), which are mapped from pixel values to a dB scale. These are combined with $P=$100 environmental features to form a 102-dimensional input vector, with the signal strengths as the label.

\textbf{Non-IID Scenario Generation:} To simulate real-world statistical heterogeneity, we partition the data into three scenarios based on signal strength variability. We define a heterogeneity metric for each location as $H = \text{std}(s_1, s_2, s_3, s_4)$. Using the 33rd and 66th percentiles of $H$ ($q_{33}, q_{66}$), we create three scenarios:
\begin{itemize}
    \item \textbf{Light Non-IID:} $H \le q_{33}$. Characterized by smooth signal transitions and low variability between base stations.
    \item \textbf{Medium Non-IID:} $q_{33} < H \le q_{66}$. Represents areas with moderate signal fluctuations and increasing data distribution disparity.
    \item \textbf{Heavy Non-IID:} $H > q_{66}$. Corresponds to complex terrains like urban canyons with sharp signal drops and highly skewed data distributions.
\end{itemize}
Within each scenario, we partition the map into a $10 \times 9$ grid to create 90 virtual clients. Each client's dataset is primarily composed of samples from its corresponding grid cell, with a small number of samples from neighboring cells, thus preserving geographical locality.

\textbf{Evaluation Metrics:}
\begin{itemize}
    \item \textbf{Accuracy:} Root Mean Square Error (RMSE) and Mean Absolute Error (MAE), reported as both micro-average (computed over all test samples) and macro-average (averaged over per-client results).
    \item \textbf{Fairness and Robustness:} Per-base-station RMSE to assess performance consistency across different propagation conditions.
    \item \textbf{Communication Cost:} Total cumulative uplink communication volume in Megabytes (MB) from the start to the end of training.
\end{itemize}

\textbf{Baseline Methods:}
\begin{itemize}
    \item \textbf{FedAvg} \cite{mcmahan2017communication}: The standard federated averaging algorithm, training a single global model.
    \item \textbf{FedFly} \cite{varyam2022towards}: An edge FL framework designed for mobile environments, representing a standard edge protocol.
    \item \textbf{PFedgb} \cite{dmitriev2021pfedgp}: A PFL method based on Gaussian processes that uses a shared deep kernel for personalization.
    \item \textbf{EPFL-REMNet (Ours)}: The proposed framework combining personalization and communication efficiency.
\end{itemize}

\subsection{Main Results: Performance Across Heterogeneity Levels}
Table \ref{tab:main_results} summarizes the core performance metrics for all methods across the three Non-IID scenarios. Several clear trends emerge from the data.

\textbf{Impact of Non-IID and Necessity of Personalization:} As the degree of Non-IID increases from light to heavy, the performance of non-personalized methods like FedAvg deteriorates significantly, with its macro RMSE reaching 6.27 in the heavy scenario. This strongly validates that for tasks with high geographical correlation, such as constructing a radio environment digital twin, personalization is not an option but a necessity to accurately capture local environmental nuances.

\textbf{Effectiveness of EPFL-REMNet's Personalization Strategy:} Among personalized methods, EPFL-REMNet substantially outperforms PFedgb. In the heavy scenario, EPFL-REMNet's macro RMSE (2.02) is approximately 64\% lower than that of PFedgb (5.63). This indicates that the "shared backbone + private head" architectural split is a more effective personalization strategy for this task than sharing a non-parametric kernel.

\textbf{Overall Superiority of EPFL-REMNet:} The EPFL-REMNet framework achieves the highest digital twin fidelity (i.e., mapping accuracy) in all scenarios, with the performance gap widening as heterogeneity increases. In the medium scenario, its macro RMSE (1.33) is 66\% lower than that of the next best method, FedAvg (3.96). Crucially, it achieves this superior fidelity with a communication cost far below that of FedAvg and FedFly, demonstrating an exceptional fidelity-communication trade-off.

\subsection{Convergence and Communication Dynamics}
\begin{figure*}[t]
  \centering

  \begin{minipage}[t]{.28\textwidth}
    \centering
    \includegraphics[width=\linewidth]{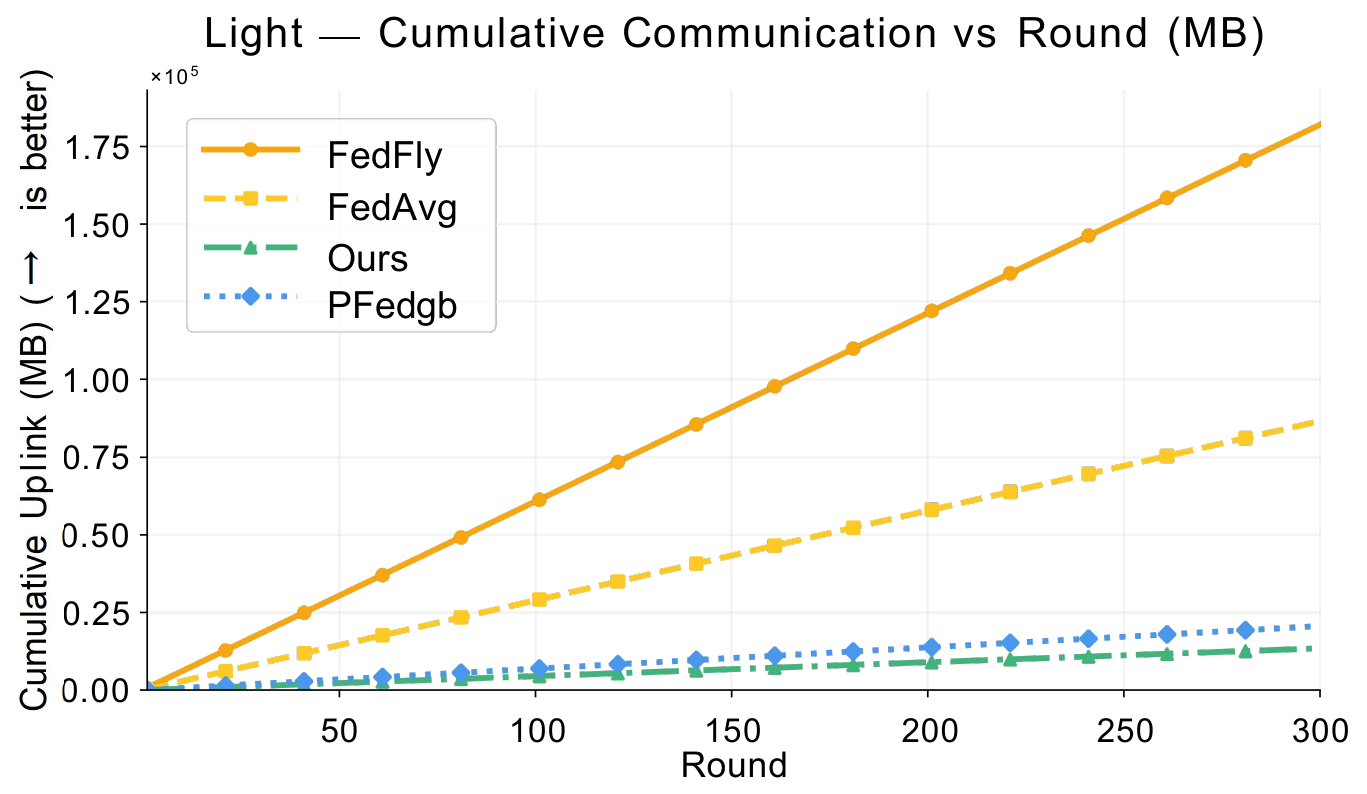}
  \end{minipage}\hfill
  \begin{minipage}[t]{.28\textwidth}
    \centering
    \includegraphics[width=\linewidth]{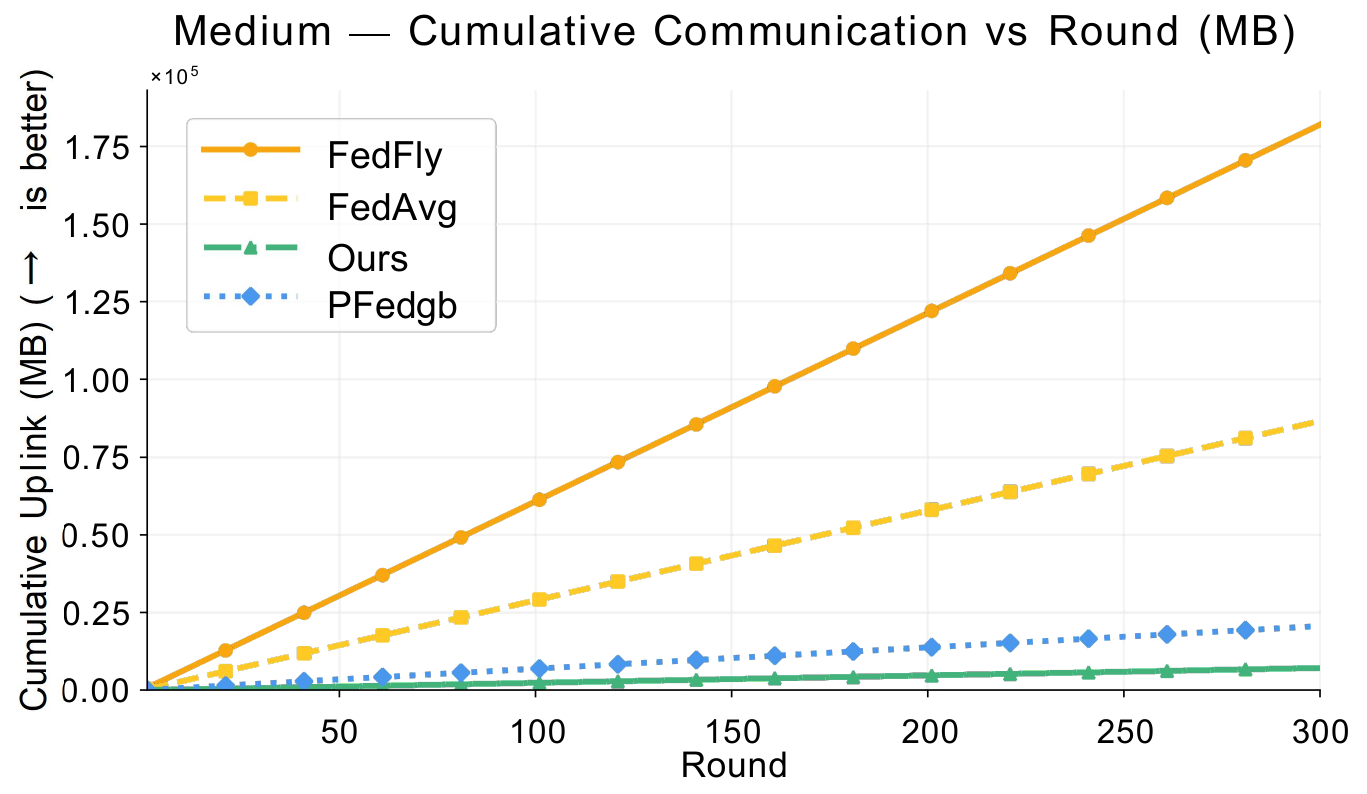}
  \end{minipage}\hfill
  \begin{minipage}[t]{.28\textwidth}
    \centering
    \includegraphics[width=\linewidth]{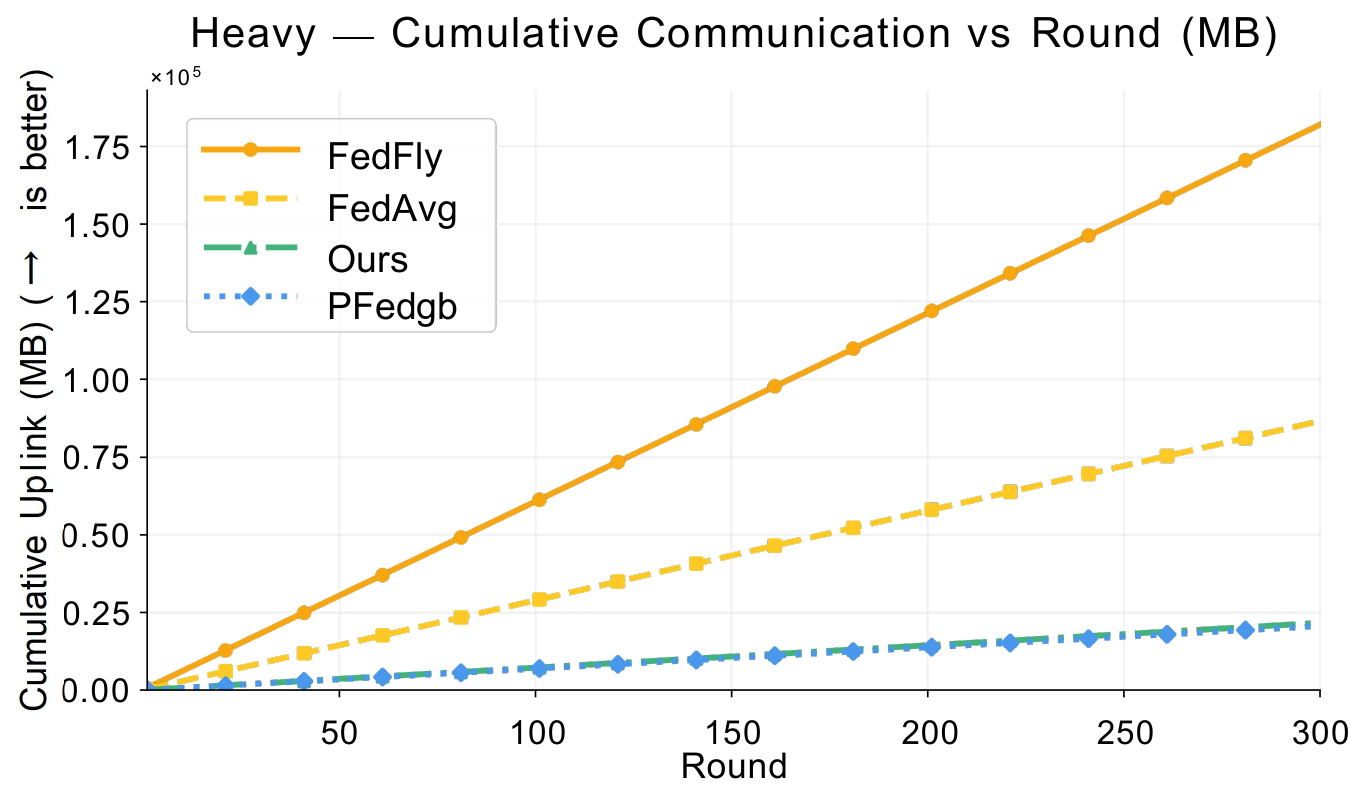}
  \end{minipage}

  \vspace{2mm}

  \begin{minipage}[t]{.28\textwidth}
    \centering
    \includegraphics[width=\linewidth]{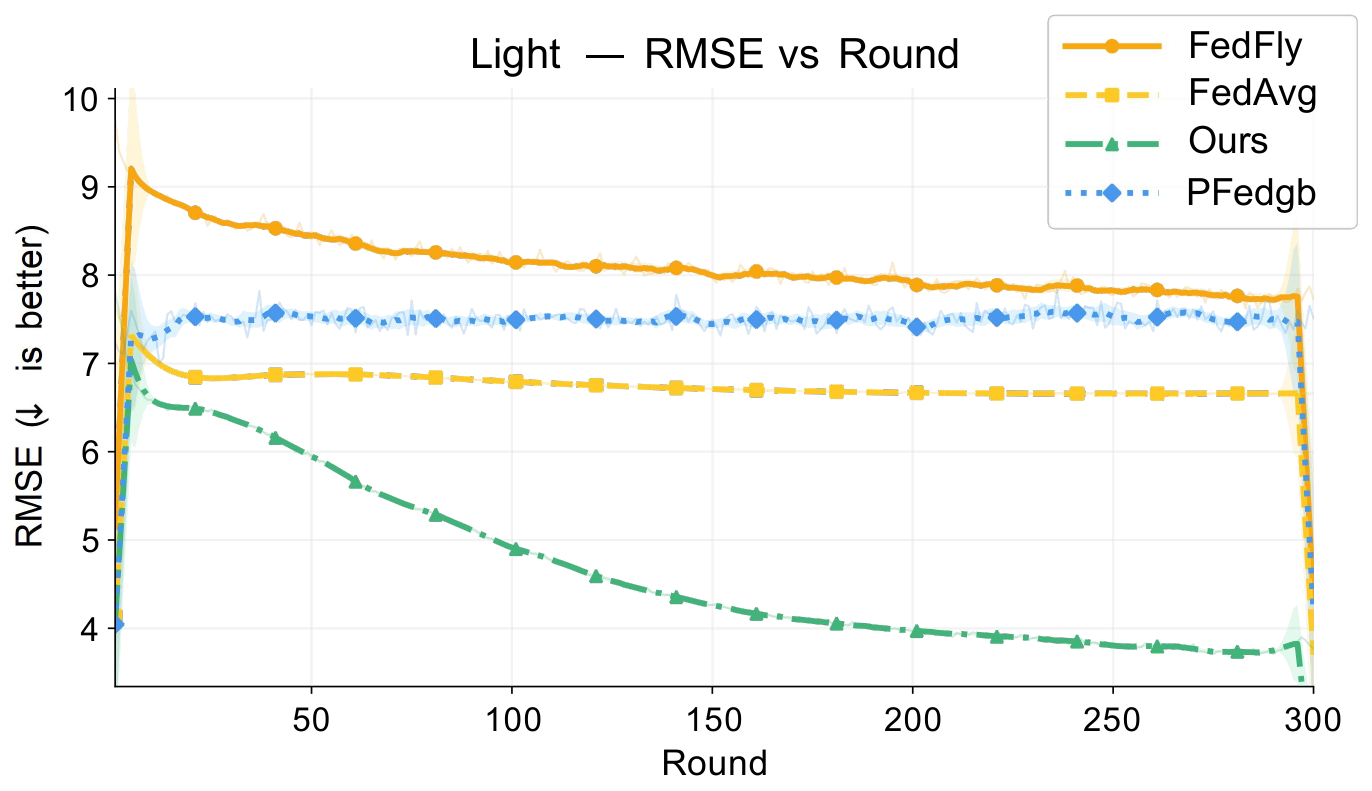}
  \end{minipage}\hfill
  \begin{minipage}[t]{.28\textwidth}
    \centering
    \includegraphics[width=\linewidth]{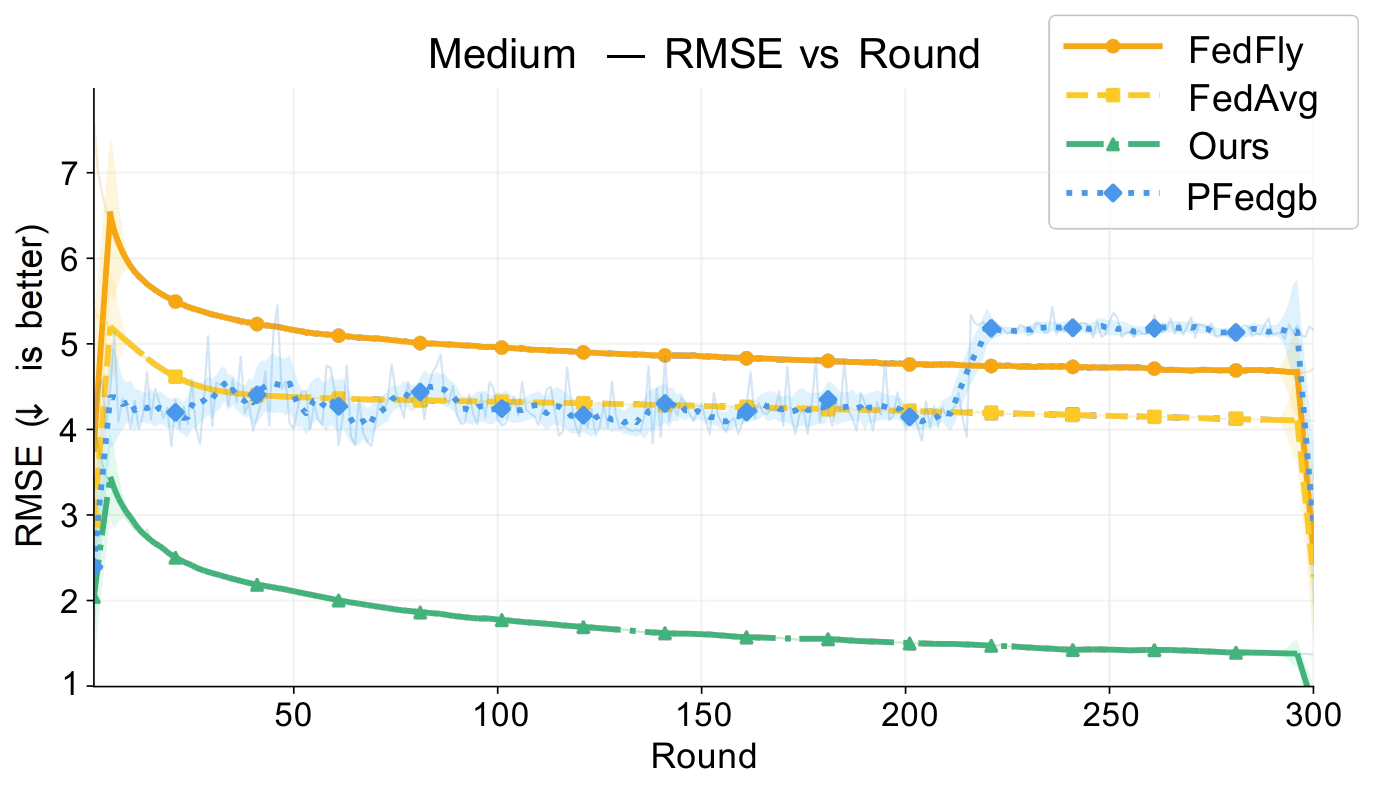}
  \end{minipage}\hfill
  \begin{minipage}[t]{.28\textwidth}
    \centering
    \includegraphics[width=\linewidth]{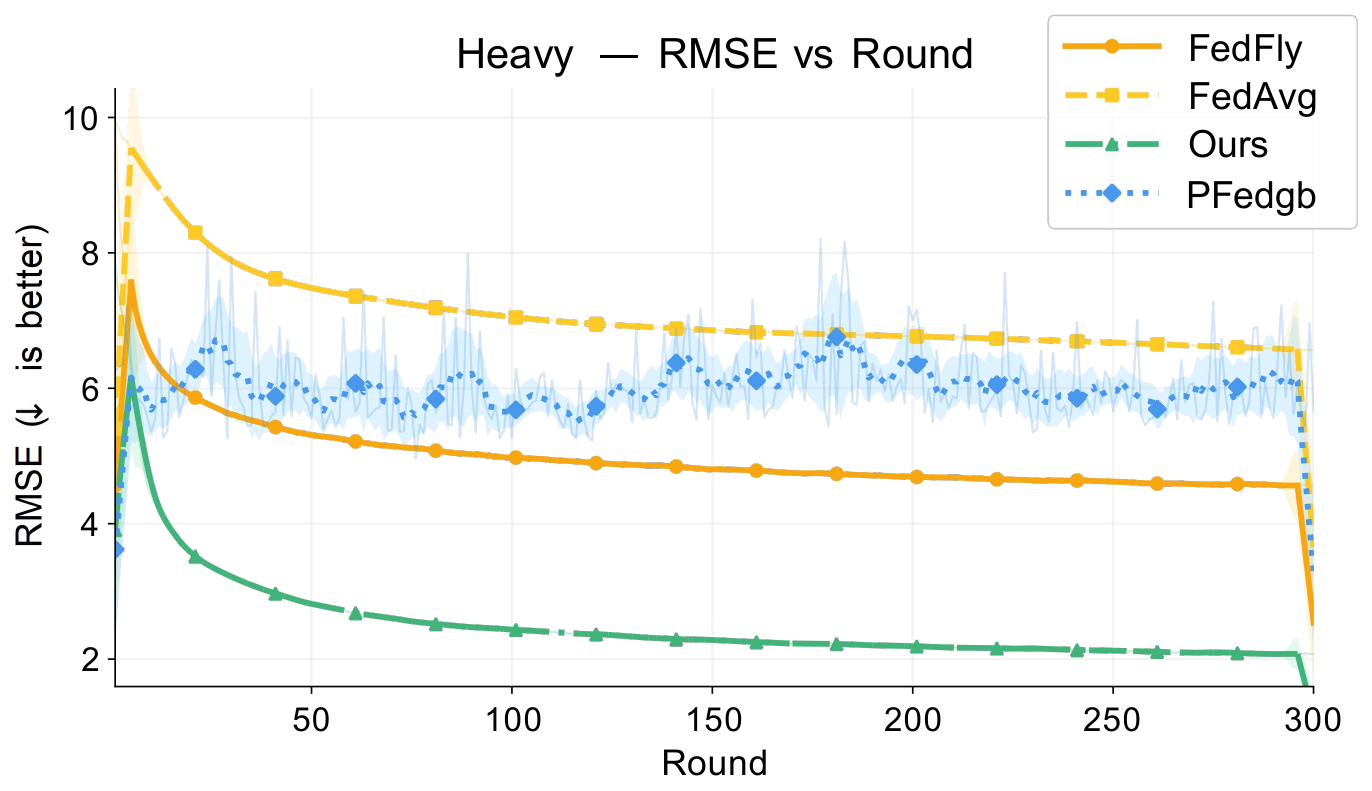}
  \end{minipage}

  \caption{Convergence curves and cumulative communication volume curves of each method in three-level Non-IID scenarios.}
  \label{fig:conv_3x2}
  \vspace{-0.5cm}
\end{figure*}

Fig. \ref{fig:conv_3x2} would visually represent the convergence behavior and communication costs. The plots would show EPFL-REMNet's macro RMSE curve rapidly descending to a much lower error floor compared to all baselines, indicating faster convergence and higher final accuracy. Concurrently, its cumulative communication curve would exhibit the flattest slope, visually confirming the high efficiency of its communication protocol. In contrast, baselines like FedAvg would show convergence to higher error levels with steeply rising communication costs.

\subsection{Accuracy-Communication Pareto Frontier Analysis}
\begin{figure*}[t]
  \centering
  \begin{minipage}[t]{.28\textwidth}
    \centering
    \includegraphics[width=\linewidth]{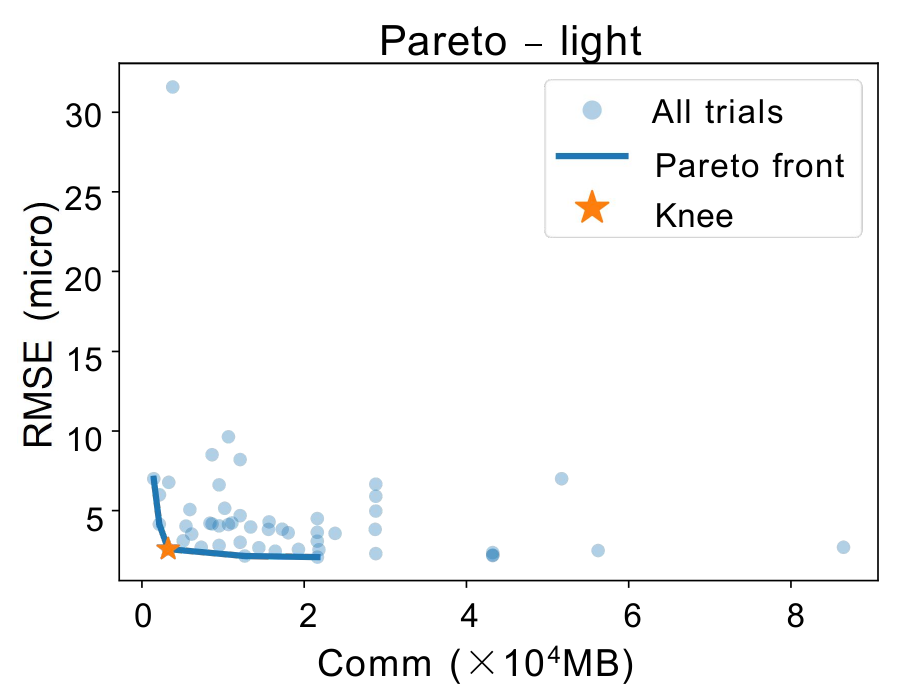}
  \end{minipage}\hfill
  \begin{minipage}[t]{.28\textwidth}
    \centering
    \includegraphics[width=\linewidth]{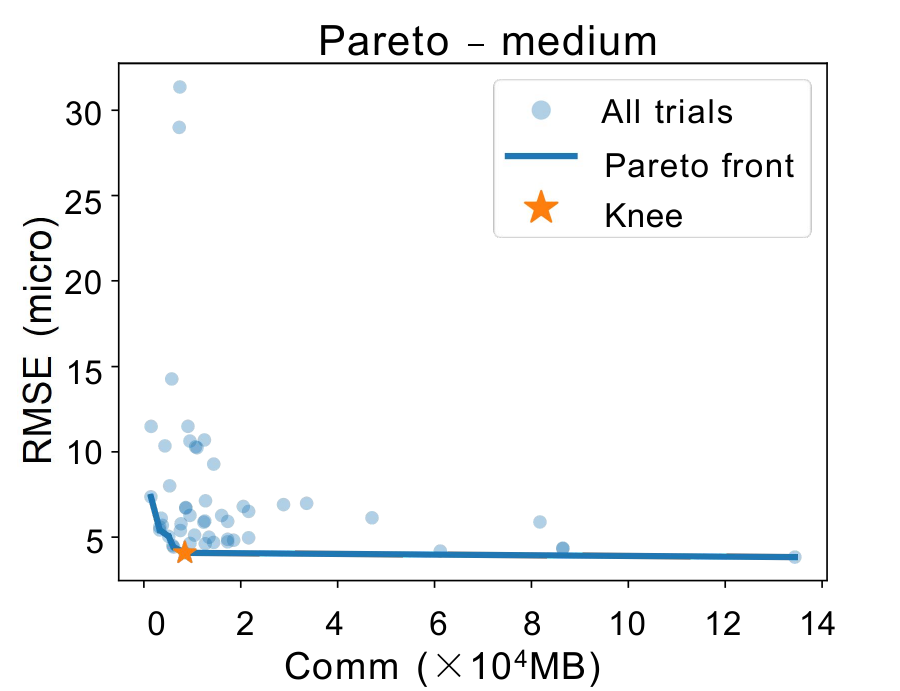}
  \end{minipage}\hfill
  \begin{minipage}[t]{.28\textwidth}
    \centering
    \includegraphics[width=\linewidth]{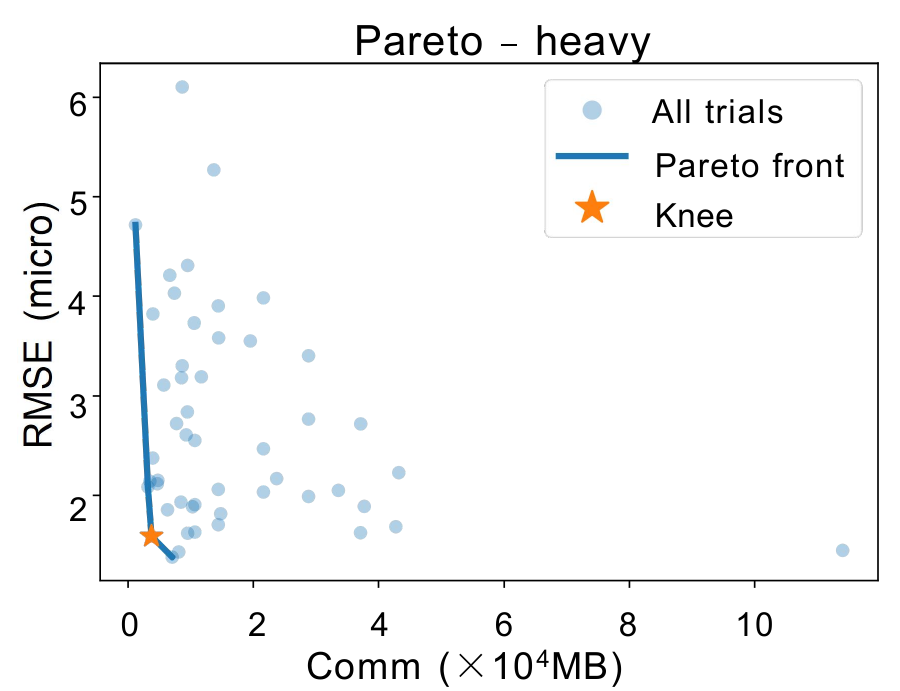}
  \end{minipage}

  \caption{Precision-communication Pareto frontier graph of EPFL-REMNet in three-level Non-IID scenarios.}
  \label{fig:pareto_front}
  \vspace{-0.6cm}
\end{figure*}

Plotting the final macro RMSE against the total communication cost for all methods (as would be done in Figure \ref{fig:pareto_front}) provides a clear visualization of the accuracy-communication trade-off. In such a plot, EPFL-REMNet would consistently occupy the bottom-left region, establishing a new and dominant Pareto frontier. This demonstrates that it offers a fundamentally better trade-off than the alternatives, achieving higher accuracy for a given communication budget or requiring less communication for a given accuracy target.


\subsection{Fairness and Robustness Analysis}
Table \ref{tab:fairness} disaggregates the RMSE by base station, providing insight into the model's robustness across different signal conditions. The data shows that while methods like FedAvg and FedFly exhibit significant performance variance across base stations and scenarios (e.g., FedFly's RMSE for base station 2 in the light scenario is 13.73), EPFL-REMNet maintains remarkably stable and low errors across all base stations and all levels of heterogeneity. The range of RMSE values for EPFL-REMNet within any given scenario is extremely narrow (e.g., 0.29 dB in the light scenario), and the variation across scenarios is also minimal. This indicates that the personalization mechanism not only improves average performance but also ensures fairness and consistency, making the system more reliable for practical deployment. Visualizing the distribution of RMSEs across all 90 clients via a box plot would further show that EPFL-REMNet has a much smaller inter-quartile range, confirming its ability to significantly uplift the performance of the worst-performing ``long-tail" clients.

\begin{table}[htbp]
\caption{RMSE (Per BS) Analysis for Fairness and Robustness.}
\label{tab:fairness}
\footnotesize
\centering
\resizebox{.4\textwidth}{!}
{\begin{tabular}{@{}llcccc@{}}
\toprule
\textbf{Scenario} & \textbf{Method} & \textbf{RMSE\_A} & \textbf{RMSE\_B} & \textbf{RMSE\_C} & \textbf{RMSE\_D} \\
\midrule
\textbf{Light} & FedAvg & 6.73 & 6.64 & 7.02 & 6.12 \\
& FedFly & 10.73 & 13.73 & 7.85 & 7.42 \\
& PFedgb & 7.54 & 7.34 & 7.68 & 6.90 \\
& \textbf{EPFL-REMNet} & \textbf{3.75} & \textbf{3.83} & \textbf{3.87} & \textbf{3.58} \\
\midrule
\textbf{Medium} & FedAvg & 4.26 & 3.50 & 4.11 & 3.95 \\
& FedFly & 4.66 & 4.66 & 4.76 & 4.75 \\
& PFedgb & 5.30 & 4.68 & 5.23 & 4.89 \\
& \textbf{EPFL-REMNet} & \textbf{1.29} & \textbf{1.24} & \textbf{1.38} & \textbf{1.39} \\
\midrule
\textbf{Heavy} & FedAvg & 6.58 & 6.11 & 6.11 & 6.26 \\
& FedFly & 4.47 & 4.64 & 4.60 & 4.56 \\
& PFedgb & 5.63 & 5.27 & 5.25 & 5.80 \\
& \textbf{EPFL-REMNet} & \textbf{2.08} & \textbf{1.77} & \textbf{2.18} & \textbf{2.04} \\
\bottomrule
\end{tabular}}
\vspace{-0.3cm}
\end{table}

\subsection{Ablation Study: Deconstructing EPFL-REMNet}
To quantify the contribution of each component in EPFL-REMNet, we conducted an ablation study in the challenging heavy Non-IID scenario. The results are presented in Table \ref{tab:ablation_results_detailed}.

\begin{table*}[htbp]
\caption{Performance Breakdown of EPFL-REMNet Variants Across Three Non-IID Scenarios.}
\label{tab:ablation_results_detailed}
\centering
\footnotesize
\resizebox{.7\textwidth}{!}{\begin{tabular}{@{}llccccc@{}}
\toprule
\textbf{Scenario} & \textbf{Variant (Method)} & \textbf{Micro RMSE $\downarrow$} & \textbf{Macro RMSE $\downarrow$} & \textbf{Comm (MB) $\downarrow$} & \textbf{$\Delta$RMSE vs Full (\%) $\downarrow$} & \textbf{$\Delta$Comm vs Full (\%) $\downarrow$} \\
\midrule
\textbf{Light} & w/o split-head & 5.6414 & 5.6115 & 2162.4 & 312.17\% / 323.13\% & -69.30\% \\
& w/o periodic sync & 6.6307 & 6.5987 & 2332.8 & 384.44\% / 397.56\% & -64.04\% \\
& w/o Top-K & 3.7590 & 3.7417 & 10811.8 & 174.64\% / 182.14\% & 53.51\% \\
& w/o quantization & 5.4228 & 5.3963 & 21620.1 & 296.19\% / 307.05\% & 206.96\% \\
& w/o EMA & 5.5410 & 5.5129 & 3458.8 & 305.00\% / 315.70\% & -50.89\% \\
& \textbf{Full} & \textbf{1.3687} & \textbf{1.3262} & \textbf{7043.2} & \textbf{0.0\%} & \textbf{0.0\%} \\
\midrule
\textbf{Medium} & w/o split-head & 2.4123 & 2.3383 & 2162.4 & -36.04\% / -37.73\% & -83.90\% \\
& w/o periodic sync & 4.0384 & 3.9253 & 2332.8 & 7.07\% / 4.54\% & -82.63\% \\
& w/o Top-K & 1.5782 & 1.5362 & 10811.8 & -58.16\% / -59.09\% & -19.54\% \\
& w/o quantization & 2.4224 & 2.3473 & 21620.1 & -35.77\% / -37.49\% & 60.90\% \\
& w/o EMA & 2.4751 & 2.4005 & 3458.8 & -34.38\% / -36.07\% & -74.26\% \\
& \textbf{Full} & \textbf{3.7717} & \textbf{3.7548} & \textbf{13437.6} & \textbf{0.0\%} & \textbf{0.0\%} \\
\midrule
\textbf{Heavy} & w/o split-head & 6.2863 & 6.0254 & 2332.8 & 203.27\% / 198.60\% & -89.22\% \\
& w/o periodic sync & 2.4999 & 2.4311 & 10811.8 & 20.61\% / 20.48\% & -50.02\% \\
& w/o Top-K & 4.5657 & 4.4361 & 21620.1 & 120.27\% / 119.84\% & -0.05\% \\
& w/o quantization & 4.5772 & 4.4326 & 3458.8 & 120.82\% / 119.66\% & -84.01\% \\
& w/o EMA & 2.7651 & 2.6893 & 2162.4 & 33.40\% / 33.27\% & -90.00\% \\
& \textbf{Full} & \textbf{2.0728} & \textbf{2.0179} & \textbf{21630.7} & \textbf{0.0\%} & \textbf{0.0\%} \\
\bottomrule
\end{tabular}}
\end{table*}

The results reveal several key findings. First, removing the personalized heads (w/o split-head) is the most detrimental change, causing the macro RMSE to skyrocket by nearly 200\%. This confirms that personalization is the foundational and most critical element for tackling data heterogeneity.

A second, and highly significant, finding is the catastrophic performance degradation observed when removing the communication compression components. Disabling Top-K sparsification or 8-bit quantization increases the macro RMSE by approximately 120\%. This counter-intuitive result suggests that these compression techniques do more than just save bandwidth; they play a crucial, positive role in the learning process itself, a point we will analyze further in the Discussion section. Finally, removing periodic synchronization (i.e., communicating more frequently) and EMA smoothing also leads to notable performance drops (20-33\%), underscoring their importance for stabilizing training. The study confirms that the components of EPFL-REMNet are complementary and their synergy is key to achieving the final ``high-accuracy, low-communication" goal.

%% file: conclusion.tex
\section{Conclusion}
\label{sec:con}
This paper proposed EPFL-REMNet, an efficient personalized federate framework for 6G heterogeneous radio environment digital twin. By synergistically co-designing a ``shared backbone + local personalized heads” architecture with novel communication efficiency pipelines—including Top-K sparsification, error feedback, quantization, and periodic updates, our framework effectively addresses the dual challenges of strong Non-IID data and constrained uplink bandwidth, which makes the digital twin equitable and robust across the entire 6G network. Our results highlight the necessity of co-optimizing personalization and communication efficiency and confirm that aggregating only the backbone updates is an effective strategy for maintaining training stability under high compression ratios.


